# Industrial Computing Systems:
# A Case Study of Fault Tolerance Analysis

Andrey A. Shchurov

*Department of Telecommunications Engineering, Faculty of Electrical Engineering,
Czech Technical University in Prague, The Czech Republic*

*Abstract—* **Fault tolerance is a key factor of industrial computing systems design. But in practical terms, these systems, like every commercial product, are under great financial constraints and they have to remain in operational state as long as possible due to their commercial attractiveness. This work provides an analysis of the instantaneous failure rate of these systems at the end of their life-time period. On the basis of this analysis, we determine the effect of a critical increase in the system failure rate and the basic condition of its existence. The next step determines the maintenance scheduling which can help to avoid this effect and to extend the system life-time in fault-tolerant mode.**

## I. INTRODUCTION

Nowadays, manufacturing companies are seeking to continuously improve efficiency and drive down costs for existing facilities and processes. The key to achieving these goals is uninterrupted access to information. With a constant flow of data (including real-time technological processes), manufacturers can develop more efficient ways to connect globally with suppliers, employees and partners, and to more effectively meet the needs of their customers. As a consequence, in addition to the technical specifications (performance, interoperability, functionality, etc.), industrial computing systems face the following additional challenges:

- reliability – solutions must support the operational availability of the manufacturing facility;
- cost – capital comes at a premium, and additional costs (or costlier components) must add clear value that is understood by the financial management;
- flexibility – solutions have to rely on commercial off-the-shelf (COTS) equipment, provided by a number of vendors.

Operational availability is the critical feature of industrial computing systems. For this reason the design of these systems is based on the concepts of fault tolerance – in practical terms, they are able to keep working to a level of satisfaction in the presence of technical and/or organizational problems, including [1]:

- hardware-related faults;
- software bugs and errors;
- physical damage or other flaws introduced into the system from the environment;
- operator errors, such as erroneous keystrokes, bad command sequences, or installing unexpected software.

The key factor of the fault tolerant design is preventing failures due to system components and it addresses the fundamental characteristic of fault tolerance in two ways [2][3]:

- replication – providing multiple identical instances of the same component and choosing the correct result on the basis of a quorum (voting);
- redundancy – providing multiple identical instances of the same component and switching to one of the remaining instances in case of a failure (failover).

On the other hand, it is well known that the effectiveness of computing systems depend on both the quality of its design as well as the proper maintenance actions to prevent it from failing. In fact, the choice of scheduled maintenance policies which are optimum from an economic point of view constitutes a predominating approach in reliability theory [4].

Our main goal is finding the simplest and cheapest solution to keep fault tolerant industrial computing systems in operational state as long as possible due to their commercial attractiveness. Thus, to accomplish such a goal we need: (1) to identify a typical (commercial) configuration of these systems; and (2) to analyse systems behaviour at the end of the useful period and at the wear-out period of the systems life-time.

The rest of this paper is structured as follows. Section 2 introduces the related work. Section 3 presents analysis of the instantaneous failure rate of commercial computing systems at the end of their life-time period. On the basis of this analysis, we determine the "Red zone" (a critical increase in the system failure rate) and the basic condition of its existence. Section 4 introduces the maintenance scheduling which can help to avoid this effect. Finally, conclusion remarks and future research directions are given in Section 5.

## II. BACKGROUND

In the past several decades, maintenance and replacement problems have been extensively studied in the literature. The most recent systematic survey of maintenance policies for the last 50 years is presented by Sarkar et al. [5]. Based on this survey, maintenance models can be roughly classified into following categories: age replacement policy, block replacement policy, periodic preventive maintenance policy, failure limit policy, sequential preventive maintenance policy, repair cost limit policy, repair time limit policy, repair number counting policy, reference time policy, mixed age policy, group maintenance policy, opportunistic maintenance policy, etc. Each kind of policy has different characteristics, advantages and disadvantages. In this context, this work lies in the area of periodic preventive maintenance policy.

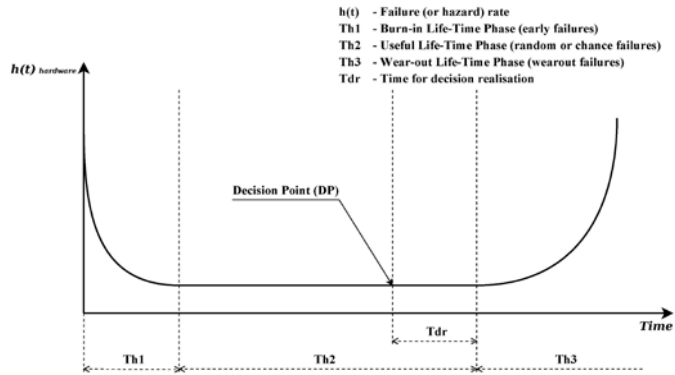

Fig. 1  Bathtub curve for electronic devices.

On the other hand, when dealing with maintenance models the analysis of the failure rate play a primary role. Generally, we can define the instantaneous failure rate as:

$$h(t)_{system} = h(t)_{hardware} + h(t)_{software} + h(t)_{operate}$$

$h(t)_{hardware}$ is hardware failure rate (defined by vendors). This is a typical bathtub curve for electronic devices (see Fig. 1) [6][7][8][9]. In this case the failure rate can be represented by the Weibull transformed distribution [7]:

$$h(t)_{hardware} = \lambda \beta t^{\beta-1}$$

if $t \in Th1$ (Burn-in Life-Time Phase)     then $0 < \beta < 1$
if $t \in Th2$ (Useful Life-Time Phase)      then $\beta = 1$
if $t \in Th3$ (Wear-Out Life-Time Phase) then $\beta > 1$

We should mention here environmental influences – temperature, humidity, EMI and other [6]. These factors exert influence not only on components/units on-the-job, but on spare components/units on-the-shelf. Bad storage conditions can directly affect hardware failure rates (the stress effect in a typical bathtub curve [7]) or even lead to unexpected failure of spare components/units in the worst case.

And based on this representation, we can define the "Decision point" (DP) – the critical point of every commercial telecommunication project – where the IT department has to decide between:
- starting a new project (buying a new system),
- buying additional spare components/units,
- finishing the current project.

$h(t)_{software}$ is embedded software failure rate (defined by vendors). The total failure rate for the software can be represented as:

$$h(t)_{software} = h(t)_{update} + h(t)_{upgrade}$$

where $h(t)_{update}$ is reliability improvement failure rate; and $h(t)_{upgrade}$ is upgrade failure rate.

The reliability improvement or bug fix software modification process occurs as part of regularly scheduled software updates. As a consequence, the reliability improvement failure rate is closely related to early failures in hardware [6][8]. In contrast to the bug fix software modifications, vendors are continuously changing embedded software to both improve existing functionality and add new capabilities. As the software grows and changes, the upgrade failure rate will inherently increase due to the increased code size and complexity. Thereafter, we have two basic options:
- Minor code changes (current software release update). In this case, the upgrade failure rate affects the total (aggregate) failure rate like the stress effect [8].
- Global code changes (upgrade to new software release). This case leads the total (aggregate) failure rate to the beginning of another burn-in period [6][8].

An important note – even the total failure rate tends to zero value as time becomes large, the processes of code changing and code size growth lead computing systems to settle on a steady-state (nonzero software failure rate) [8].

$h(t)_{operate}$ is operator failure rate – erroneous keystrokes, bad command sequences, or installing unexpected software [1].

### III. ANALYSIS OF SYSTEM BEHAVIOUR

When talking about fault tolerant industrial computing systems, we usually mean redundant commercial computing systems (we need to state here – specific areas like the military, nuclear or aerospace industries are beyond the scope this work). In practice these industrial systems are under great financial constraint – the main challenge is how to combine a real fault tolerance and commercial attractiveness. As a consequence, nowadays these systems have modular and/or distributed architectures with critical components duplication (usually controller/processor and power supply units). Additional reliability is provided by the availability of spare components or units. The number and composition are defined by the project's budget. The architectural diagram of these systems (based on the von Neumann machine representation [10]) is shown in Fig. 2. Of course, some vendors provide, within extended technical support, operative replacement of failure components, but this service has disadvantages:
- additional expenses – it is very difficult to find strong arguments for financial management;
- response time (especially in developing countries) is always longer than having a spare component on-the-shelf.

Thus, as the object for analysis we have a system with two controller units on-the-job and one spare controller unit on-the-shelf (see Fig. 2). Controller/processor units are usually the most expensive part of every computing systems and it is usually impossible to persuade the financial management to buy more than one spare unit.

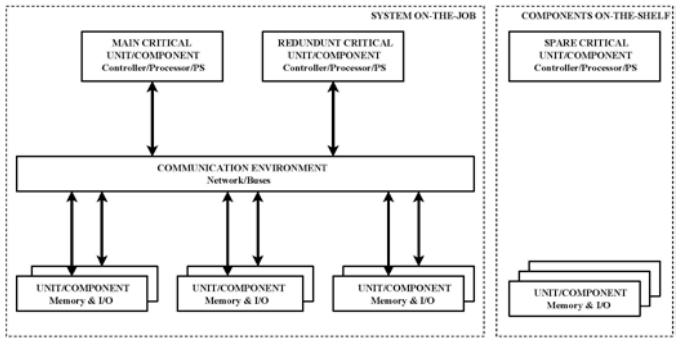

Fig. 2 Architectural diagram of redundant commercial computing systems.

The analysis covers the end of the useful period and the wear-out period of the system life-time. In order to simplify the analysis, let us make the following assumptions:
- All three controller units are identical. Components/units instantaneous failure rate is:

$$h(t)_{controller1} = h(t)_{controller2} = h(t)_{controller3} = h(t)_{controller}$$

- Two main controllers units on-the-job are used during their entire life-time periods. The spare controller unit is used only if one of the two main controllers fails (Interaction Type_1 – see Section 4 "System maintenance scheduling").
- This standby redundant system has perfect sensing and switching subsystems.
- The IT department is staffed by qualified personnel and the system is stable and does not usually require operator interventions:

$$h(t)_{operate} << h(t)_{hardware}$$

At the end of the useful period industrial computing systems generally use "stable" software releases. In this case [8]:

$$h(t)_{software} << h(t)_{hardware}$$

Thus, the reliability function is dominated by hardware failures and the impact of software failures is minor with respect to the system failure rate:

$$h(t)_{hardware} \approx h(t)_{controller}$$

In turn, the components/units life-time period can be described by the lognormal distribution [7]. The parameters of the distribution:
- a mean $\mu$ – a mean value of components/units life-time;
- a standard deviation $\delta$ – spread of components/units life-time

Therefor the system instantaneous failure rate can be represented as:

$$h(t)_{system} = F(h(t)_{controller1}, h(t)_{controller2})$$
$$h(t)_{controller1} \approx h(t)_{hardware} = f(t)$$
$$h(t)_{controller1} \approx h(t)_{hardware} = f(t + \delta)$$
$$f(t) = \lambda\beta t^{\beta-1}$$

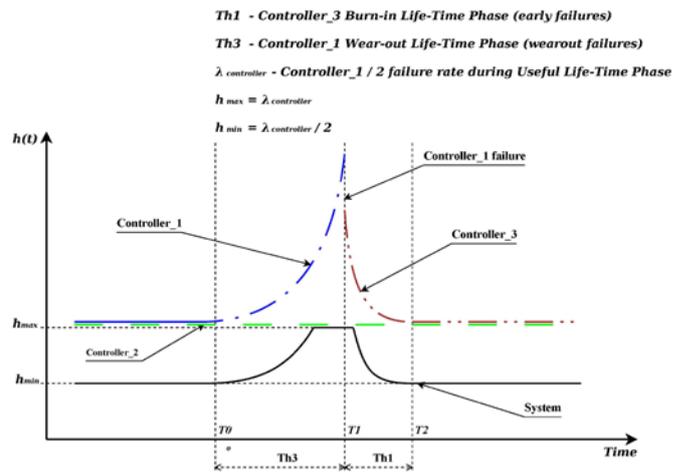

Fig. 3 Fault-tolerant system behaviour – an arbitrary component/unit failure.

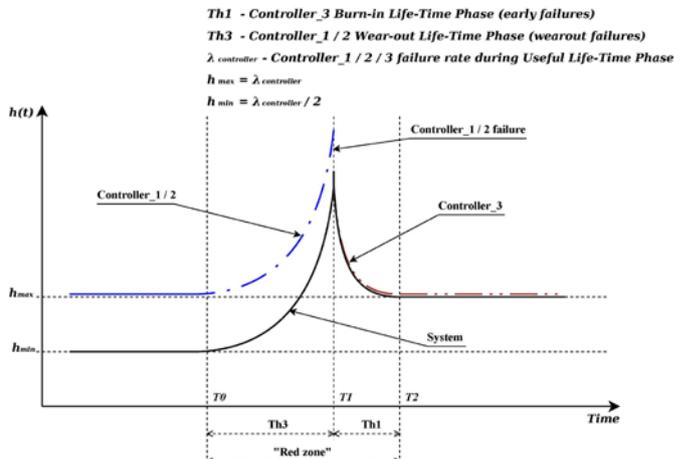

Fig. 4 Fault-tolerant system behaviour – two components/units simultaneous failure

The following two options describe various scenarios of the fault-tolerant system behaviour.

*A. Option 1 – system behaviour in the case of $\delta >> 0$*

In practical terms, this option is the current practice (failed components/units replacement) and there is nothing new here [7][8][9] – see Fig. 3.

*B. Option 2 – system behaviour in the case of $\delta \to 0$*

1. If $t < T0$ (see Fig. 4), then:
   - The first controller unit (Controller_1) is in the Useful Life-Time Phase.
   - The second controller unit (Controller_2) is in the Useful Life-Time Phase.
   - The third (spare) controller unit (Controller_3) is not present.

Thus
$$h(t)_{controller1} = h(t)_{controller2} = \lambda, \ \beta = 1$$

And
$$h(t)_{system} = F(h(t)_{controller1}, h(t)_{controller2}) = \lambda/2$$

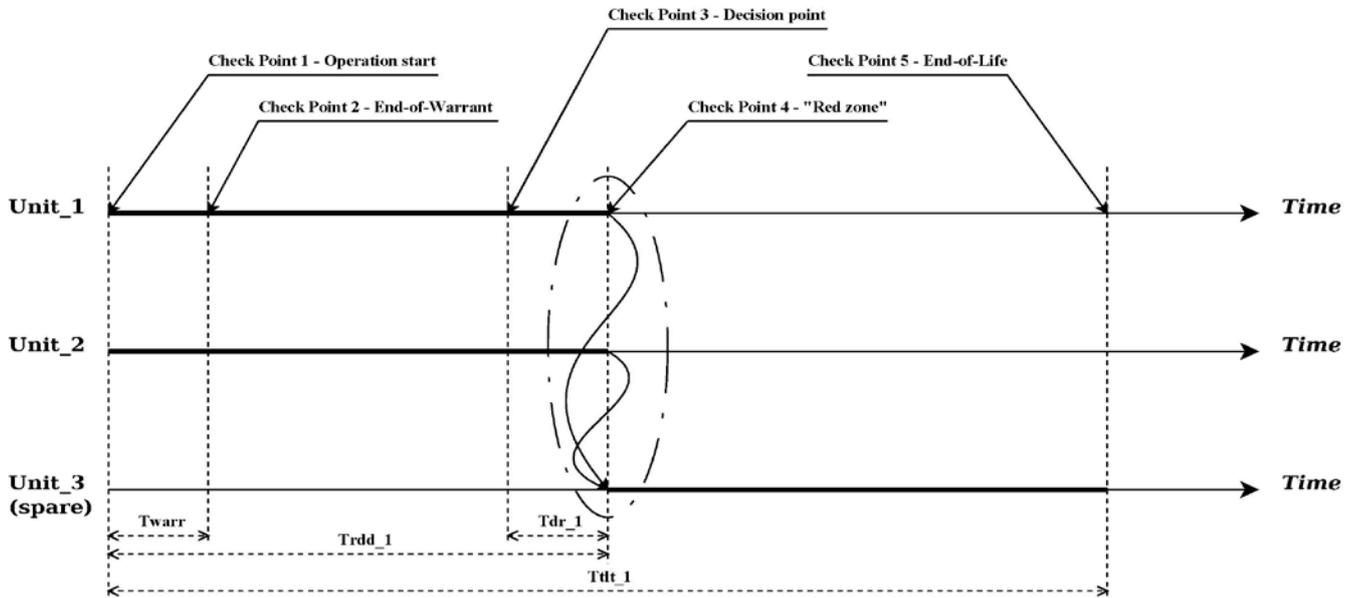

Fig. 5 Time diagram of controllers' life-time usages – Interaction Type_1.

2. If $T0 < t < T1$ (see Fig. 4), then:
- The first controller unit (Controller_1) is in the Wear-Out Life-Time Phase.
- The second controller unit (Controller_2) is in the Wear-Out Life-Time Phase.
- The third (spare) controller unit (Controller_3) is not present.

Thus
$$h(t)_{controller1} = \lambda\beta t^{\beta-1}, \quad \beta > 1$$
$$h(t)_{controller2} = \lambda\beta (t + \delta)^{\beta-1}, \quad \beta > 1$$

And
$$h(t)_{system} = F(h(t)_{controller1}, h(t)_{controller2})$$

3. If $T1 < t < T2$ (see Fig. 4), then:
- The first controller unit (Controller_1) is not present.
- The second controller unit (Controller_2) is not present.
- The third (spare) controller unit (Controller_3) is in the Burn-in Life-Time Phase.

We need to state here: the well-known practice is to burn-in components in the lab before putting them on-the-shelf – it can help to avoid the worst effect of the Burn-in Life-Time Phase. But these lab tests usually last one or two weeks (up to four in the best case) while a typical Burn-in Life-Time Phase is about 20 weeks [9]. Therefore we cannot completely eliminate this period from the analysis.

Thus
$$h(t)_{controller3} = \lambda\beta t^{\beta-1}, \quad 0 < \beta < 1$$

And
$$h(t)_{system} = F(h(t)_{controller3})$$

4. If $t > T2$ (see Fig. 4), then:
- The first controller unit (Controller_1) is not present.
- The second controller unit (Controller_2) is not present.
- The third (spare) controller unit (Controller_3) is in the Useful Life-Time Phase.

Thus
$$h(t)_{controller3} = \lambda, \quad \beta = 1$$

And
$$h(t)_{system} = F(h(t)_{controller3}) = \lambda$$

Modern industrial technologies provide an effective improvement in the stability of production processes. In turn, this fact leads to the repeatability of the technical characteristic (at least within the same production lot). And as a consequence, we have components/units with a very small spread in the components/units life-time ($\delta \to 0$). Thus, both main controllers units on-the-job come up to Wear-out Life-Time Phase almost simultaneously (with a very small spread). But at the same time a spare controller unit on-the-shelf is still in Burn-in Life-Time Phase. Therefore, we have a critical increase in the system failure probability – the "Red zone" – Fig. 3. The basic condition of the "Red zone" existence is the parameter ratio:

$$\delta < Th3$$

where $\delta$ is the spread of components/units life-time; and $Th3$ is the duration of Wear-Out Life-Time Phase (see Fig. 4).

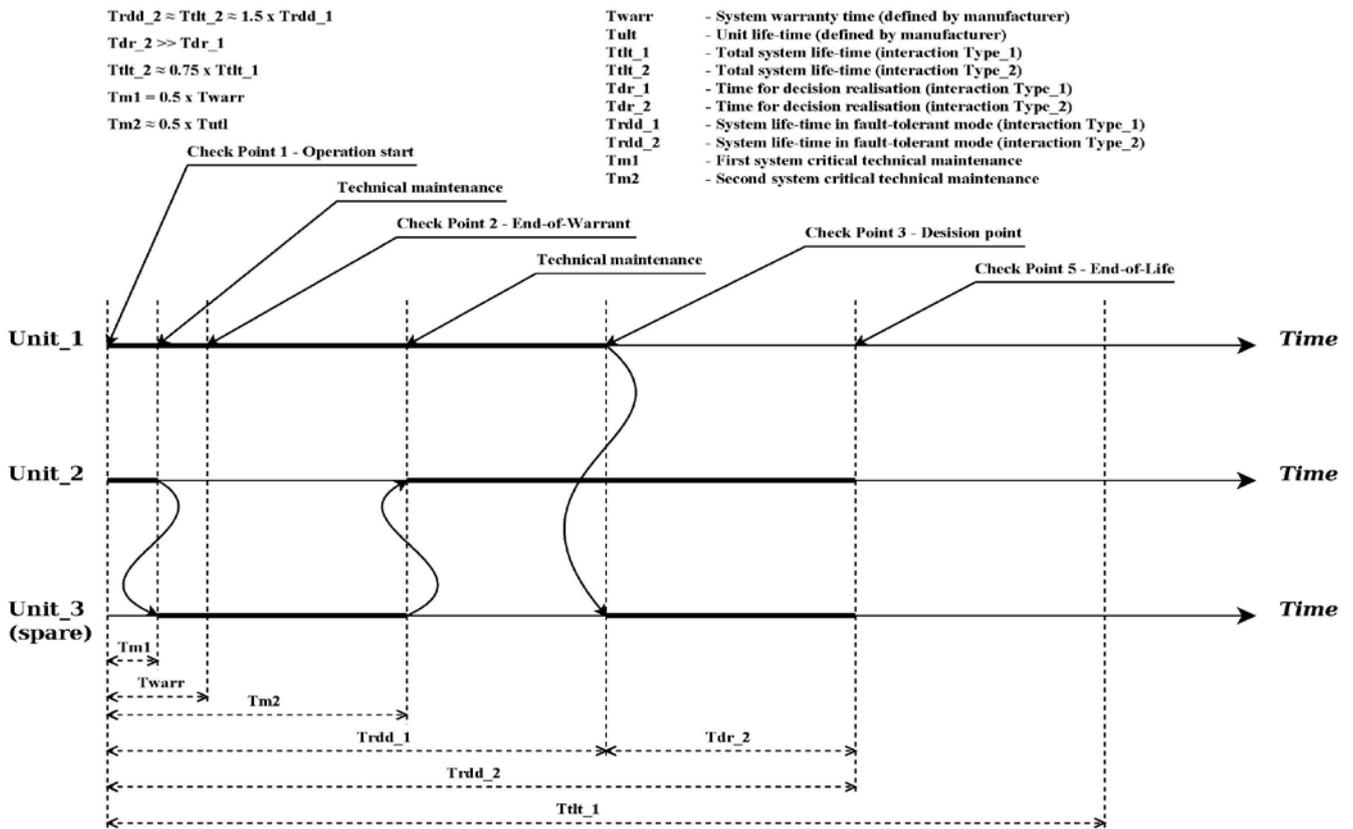

Fig. 6 Time diagram of controllers' life-time usages – Interaction Type_2.

This effect can cause considerable problems for IT departments. And in this case the fault tolerant (redundant) design only cannot protect against it.

## IV. SYSTEM MAINTENANCE SCHEDULING

The previous section presents the formal description of the effect of a critical increase in the system failure rate. And now our main goal is finding the simplest and cheapest solution to avoid this effect for existing systems. It is obvious, being under continuous financial constraints, that managerial procedures (maintenance policy) are the most appropriate way.

Again our system has two controller units on-the-job and one spare controller unit on-the-shelf (see Fig. 2). Fig. 5 and 6 show the time diagrams of controller units' life-time usages.

### A. Interaction Type_1

Fig. 5 presents the "classical" approach – two main controller units on-the-job are used for the whole of their life-time periods. The spare controller unit is used iff one of the two main controllers fails.

Interaction Type_1 characteristic features:
- In this case we have the potential condition for the "Red zone" existence.
- It is very difficult to determine DP correctly – we can use only vendors' statistics (MTBF) and in the real world statistics very often lie. But a mistake in DP determination carries reputation risks for IT department personnel:
  - too early assessment – in this case an IT department will very probably have problems from financial management (unnecessary investment);
  - too late assessment – in this case it is highly probably that the system will reach the wear-out period (the "Red zone" in the worst case) and only the IT-department (not financial management) takes full responsibility for the consequences.
- In this case it is very difficult to convince financial management of the need for investment in IT infrastructure – the system has been working well since installation and there are spare critical components/units on-the shelf.

But we need to state here – the real advantage of this case is the minimal IT department interference in error-free system operations.

### B. Interaction Type_2

Fig. 6 presents the possible solution based on periodic replacement of one of two main controller units and a spare controller unit.

Interaction Type_2 characteristic features:
- In this case we do not have the potential condition for the practical "Red zone".

- It is very easy to determine DP – the system is still in fault-tolerant mode but there are no longer any spare critical components/units (see Fig. 6). And it is obvious that in this case we have a lot of time for the decision realization (starting a new project or buying additional spare components/units).
- In this case there is the strong argument for financial management – there is nothing on-the-shelf.

And we need to state here – in this case the system life-time in fault-tolerant (redundant) mode is up to 50% longer than the system life-time in the first case (Interaction Type_1). Potentially it can be used for saving investments in IT infrastructures.

## V. CONCLUSIONS

When talking about fault tolerant industrial computing systems, we usually mean redundant commercial computing systems (specific areas like the military, nuclear or aerospace industries are beyond the scope this work). In practice these industrial systems are under great financial constraint. As a consequence, they have to remain in operational state as long as possible due to their commercial attractiveness.

In this work we provided the analysis of the instantaneous failure rate of commercial redundant computing systems at the end of their life-time period. Under certain circumstances the repeatability of the technical characteristic can cause a critical increase in the system failure rate for redundant systems at that time. The fault tolerant (redundant) design cannot protect against this challenge (in contrast to The Useful Life-Time Phase). In this case, the significant impact on operational availability characteristics can be provided by the maintenance scheduling. On the basis of the analysis we determined the maintenance scheduling which can help (1) to avoid this effect; and, as a consequence, (2) to extend the system life-time in fault-tolerant (redundant) mode.

### ACKNOWLEDGMENT

This research has been performed within the scientific activities at the Department of Telecommunication Engineering of the Czech Technical University in Prague, Faculty of Electrical Engineering.